\documentclass{article} % For LaTeX2e
\usepackage{colm2024_conference}

\usepackage{microtype}
\usepackage{hyperref}
\usepackage{url}
\usepackage{booktabs}
\usepackage{graphicx}
\usepackage{caption}
\usepackage{listings}
\usepackage{comment}
\usepackage{subcaption}

\colmfinalcopy 

\title{Building a Domain-specific Guardrail Model in Production}
\author{Mohammad Niknazar$^{1}$,
Paul V Haley$^{1}$, 
Latha Ramanan$^{1}$,
Sang T. Truong$^{2}$, \And
Yedendra Shrinivasan$^{1}$,
Ayan Kumar Bhowmick$^{1}$,
Prasenjit Dey$^{1}$, \And
Ashish Jagmohan$^{1}$,
Hema Maheshwari$^{1}$,
Shom Ponoth$^{1}$,
Robert Smith$^{1}$, \And
Aditya Vempaty$^{1}$,
Nick Haber$^{2}$,
Sanmi Koyejo$^{2}$,
Sharad Sundararajan$^{1}$  \\
  \\ $^{1}$ Emergence AI
  \\ $^{2}$ Stanford University
}
\date{March 2024}

\begin{document}
\maketitle

\begin{abstract}
Generative AI holds the promise of enabling a range of sought-after capabilities and revolutionizing workflows in various consumer and enterprise verticals. However, putting a model in production involves much more than just generating an output. It involves ensuring the model is reliable, safe, performant and also adheres to the policy of operation in a particular domain. Guardrails as a necessity for models has evolved around the need to enforce appropriate behavior of models, especially when they are in production. In this paper, we use education as a use case, given its stringent requirements of the appropriateness of content in the domain, to demonstrate how a guardrail model can be trained and deployed in production. Specifically, we describe our experience in building a production-grade guardrail model for a K-12 educational platform. We begin by formulating the requirements for deployment to this sensitive domain. We then describe the training and benchmarking of our domain-specific guardrail model, which outperforms competing open- and closed- instruction-tuned models of similar and larger size, on proprietary education-related benchmarks and public benchmarks related to general aspects of safety. Finally, we detail the choices we made on architecture and the optimizations for deploying this service in production; these range across the stack from the hardware infrastructure to the serving layer to language model inference optimizations. We hope this paper will be instructive to other practitioners looking to create production-grade domain-specific services based on generative AI and large language models.
\end{abstract}

\section{Introduction} \label{sec_introduction}

The advanced capabilities of the latest Large Language Models (LLMs) in generating and interpreting highly coherent, human-like text unleash significant potential for diverse applications, including content creation for marketing, customer service chatbots, education and e-learning, medical assistance, finance, and legal support. However, deploying LLM-based applications carries inherent risks. There have been numerous incidents where LLM-based applications have erroneously enabled particular policies for the purchase of products, used offensive language, disseminated incorrect information, or even provided guidance on unethical activities such as suggestions for the best choice of arms for a particular type of activity. These risks underscore the critical need for robust safety and reliability measures, especially when LLM-based applications are used in production, be it in consumer domains or enterprise. Thus, developing LLM-based applications demands a tradeoff in harnessing the general linguistic abilities of LLMs while simultaneously ensuring they strictly adhere to the specified behavior required for a particular application. 

Guardrails could be internal to LLM which means it has been trained and aligned to adhere to a particular policy, or it could be external where external rules or mechanisms can be applied to the input query to decide whether to proceed with the query and in what manner, and also monitor the output of the LLMs to check for adherence to the policy before it is sent to the user. The key challenge in developing efficient guardrails lies in clearly defining the requirements and expectations from the model. For instance, regulations differ by industry, country, and region. Additionally, ethical considerations such as fairness or the avoidance of offensive responses are challenging to concretely and actionably specify (\cite{dong2024buildingguardrails}, \cite{suriya2024stateofguardrails}). Guardrails can be categorized primarily into the following types:

\begin{itemize}
    \item {\bf Domain Specific Guardrails}: This set of guardrails deals with ensuring adherence of the output of the model to a particular context or domain. For example, in finance, the meaning and implications of "securities" is completely different from that in the IT operations domain.
    \item {\bf Legal/Compliance Guardrails}: Different domains have different compliance requirements and hence some actions or outputs are not allowed. For example in the healthcare domain, HIPPA disallows any release of personally identifiable information, or in education FERPA requires that no student records can be released to anyone without the consent of parents for children below the age of 18.  
    \item {\bf Ethical Guardrails}: This guardrail deals with the general human and societal implications of the actions of a model. This includes aspects such as fairness, transparency, privacy etc.
    \item {\bf Safety and Security Guardrails}: This aspect of the guardrail aims to prevent harm and use of the model for wrongful purposes. This includes changing the behavior of the model through prompt injection, jailbreaking, as well as use of the model to perform malicious actions using different tools.
\end{itemize}

The metrics and nuances of guardrails in different domains are being actively studied, especially the ones that are highly regulated such as finance (\cite{narayanan2024llmrisk}), healthcare (\cite{daniel2024medicalguardrails} and education. In this paper, we examine the issues and share our experiences of building a guardrail model in the context of the education domain where guardrails are very important and requirements are fairly stringent. Implementing a real-time, production-grade guardrail LLM for the education domain presents its own significant set of challenges. This is because educational LLMs must meet unique needs such as 1) Complying with data privacy regulations like  \href{https://www2.ed.gov/policy/gen/guid/fpco/ferpa/index.html}{FERPA} and \href{https://www.ftc.gov/legal-library/browse/rules/childrens-online-privacy-protection-rule-coppa}{COPPA}, 2) Ensuring the safety and appropriateness of content, and 3) Delivering real-time responses in classrooms requiring low-cost and low-latency performance. To address these challenges and ensure the successful deployment of Safety and Appropriateness models in educational AI solutions, establishing clear and measurable performance targets (known as Service Level Objectives or SLOs) is crucial. These SLOs become part of a broader agreement called a Service Level Agreement (SLA). Furthermore, school districts may require AI developers to follow State level AI Policies/Legislation guidance (\href{https://www.dpi.nc.gov/news/press-releases/2024/01/16/ncdpi-releases-guidance-use-artificial-intelligence-schools}{NCDPI releases guidance on the use of artificial intelligence in schools | NC DPI}) to determine whether a given AI solution tool is safe and reliable to deploy in their infrastructure.  

%\paragraph{Abridged Related work} 
Recent efforts to develop LLMs for generating human-like questions for educational assessments include (\cite{wang2022towards, elkins2023useful, bulathwela2023scalable}). Several attempts have also been made to use ChatGPT for generating educational content through prompt engineering \cite{adeshola2023opportunities, baidoo2023education} with some of them focusing on generating content related to schools~\cite{jauhiainen2023generative}. However, there exist certain challenges of using models like ChatGPT for generating safe and appropriate content related to the education domain~(\cite{rahman2023chatgpt, kasneci2023chatgpt}). As a result, while progress has been made in recent literature towards developing LLMs for the education domain, there is still a dearth of research efforts ensuring safety and appropriateness while building educational LLMs. %A longer survey of related work is presented in Appendix \ref{sec:relatedworkappendix}.

There have also been attempts made in existing literature focusing on using generative AI for building production-grade domain-specific LLMs. However, one primary limitation of all these endeavors lies in the scalability and computational requirements of training and finetuning LLMs in such scenarios. Building production-grade domain-specific LLMs often necessitates vast computing resources and data, which may pose significant challenges for organizations with limited resources or infrastructure. The quality and diversity of training data can be another limitation, particularly for niche or specialized domains where annotated data may be scarce or biased, leading to suboptimal model performance and generalization. Moreover, the interpretability and explainability of domain-specific LLMs remain significant challenges, which is crucial for building trust and accountability, especially in sensitive domains like education. Data privacy and security concerns may arise when deploying domain-specific LLMs, as these models may inadvertently leak sensitive information or be susceptible to adversarial attacks. 

In response to the challenges listed above, we propose SPADE (Safe and Performant AI Deployment with continuous Evaluation), a system for safety and appropriateness that is unique to K-12 education system and a production process aimed at optimizing the performance of our finetuned LLM system while ensuring explainability of model outputs, scalability and addressing privacy concerns. In doing so, we make the following contributions:
\begin{itemize}

\item  We formulate the requirements for a safety and appropriateness system to provide a verdict for appropriate/inappropriate input of variable text length. 

\item We present a methodology for fine-tuning LLMs to optimize them for production use. We evaluate it for safety and appropriateness in the context of education and demonstrate that it outperforms competing models on proprietary and public benchmarks. 

\item   We investigate the optimized deployment of LLM-based Safety and Appropriateness service and demonstrate the impact of various design choices. 
\end{itemize}
\begin{figure}
    \centering
    \includegraphics[width=1\linewidth]{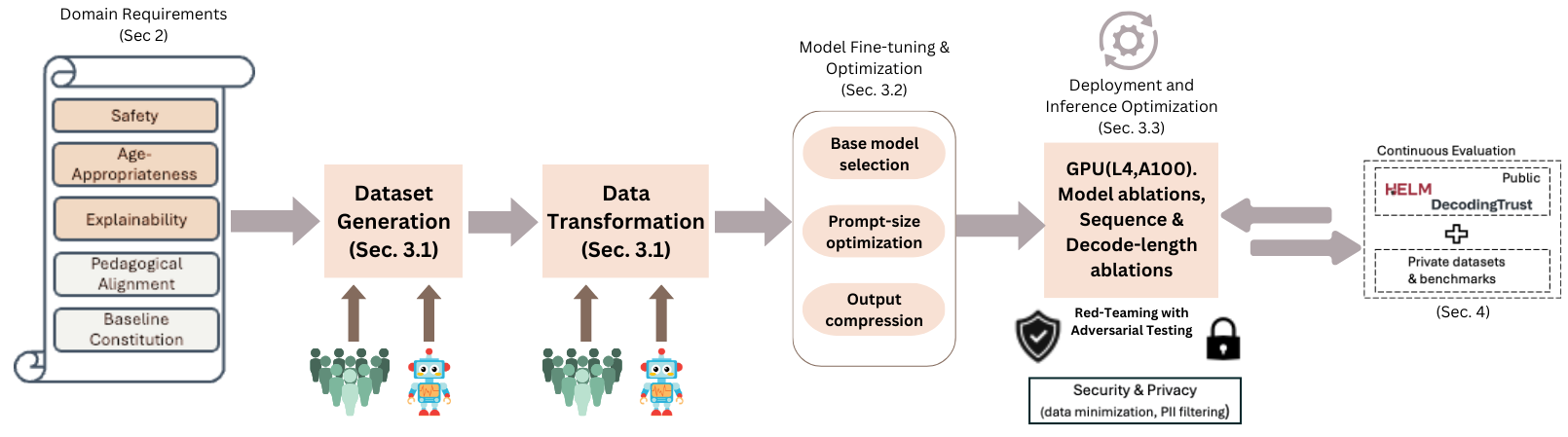}
    \caption{The SPADE system guides the lifecycle from policy and adaptation in data and model preparation through deployment, with a strong focus on continuous evaluation. SPADE ensures that the models are not only efficient and effective in real-world applications but are also trustworthy.}
    \label{fig:anchor}
\end{figure}

\section{Preliminary: Safety and Appropriateness Guardrail } \label{sec:safety}

 There are several state-of-the-art works in recent literature that focus on developing Responsible AI models with safety as the primary goal. For instance, ~\cite{madhavan2020toward} have outlined the policy considerations surrounding AI development. Recently, there has been literature on limiting general-purpose chatbots in the range of topics they can chat about according to the normative concept of appropriateness \cite{hendrikappropriateness}.

However, there was none that combined the safety and appropriateness for educational needs. Many currently deployed educational chatbots leverage wrappers around ChatGPT, catering to both educators and students within or outside the classroom environment. However, ensuring the safety and appropriateness of responses for students using these chatbots on personal devices or for educators integrating them into classroom instruction remains a critical challenge in the education domain. Inconsistency persists in how these systems define 'appropriate' and 'safe' content for educational purposes. To ensure responsible deployment of AI in K-12 education, the model needs to encompass the following key elements: 

\begin{enumerate}
    \item Prioritizes the safety requirements of the school district (for students and teachers) to prevent harmful content, such as hate speech, misinformation, bias, sexual, conspiracy theories, violence, scams, etc.  %We used Google Vertex Responsible AI dimensions for initial categorization of harmful content. 
    \item  Must also adhere to age-appropriateness by tailoring interactions to avoid complex topics and sensitive topics unsuitable for elementary-age school kids.
    \item Must have clear interpretability and explainability where applicable.
    \item Must employ curriculum-aligned content that can meet the grade-level developmental needs to be pedagogically aligned and appropriate. 
    \item Must provide a baseline constitution that helps determine the safety and appropriateness for Education. It should be customizable or configurable to contextualize the local, state, and federal requirements for safety and appropriateness.
\end{enumerate}

This paper focuses on the first 3 items. The rest are for future work.

\section{Training and Deploying Production-Grade Safety LLMs}

\subsection{Dataset Generation}

To generate a dataset for evaluating query appropriateness, we identify diverse categories of queries encountered in elementary classroom interactions with a chatbot. These categories were delineated into safe (appropriate), controversial or irrelevant (inappropriate), and unsafe (inappropriate) topics. The dataset encompassed a broad spectrum of topics, including 100 unsafe topics (e.g., pornography, racist ideologies), 50 controversial or irrelevant topics (e.g., divorce and separation), and 120 safe topics (e.g., the advantages of recycling.). 

Further data examination and enrichment added a variety of queries, including those related to explicit content, and safe personality and non-personality queries. In total, the dataset consisted of 23,304 instances. Upon reviewing the assumed labels of 'appropriate' or 'inappropriate', discrepancies in the precision of these labels were noted, leading to a comprehensive reclassification. This refined classification resulted in 14,251 'appropriate' and 9,053 'inappropriate' samples.

Text variability, including capitalization and punctuation modifications, was introduced to enhance the dataset's robustness. 25\% of the enhanced dataset was then randomly selected, maintaining the original proportion of 'appropriate' to 'inappropriate' labels, for advance annotation using a state-of-the-art LLM. This portion of the dataset was used to train the model presented in this paper.\footnote{We plan to opensource a version of this dataset for public use.} We aim to develop an advanced model that can not only generate labels but also provide scores reflecting the relevance of the input to a set of significant attributes for determining appropriateness. The attributes used for this model were adapted from the Responsible AI Vertex AI Google Cloud\footnote{\href{https://cloud.google.com/vertex-ai/docs/generative-ai/configure-safety-attributes-palm}{https://cloud.google.com/vertex-ai/docs/generative-ai/configure-safety-attributes-palm}}.

For instance, consider the following query: "Do you think that all people who practice certain religions are extremists or radicals?" The model's response was: 

\begin{lstlisting}
{"Appropriateness": "inappropriate", 
"Derogatory": 0.2, "Toxic": 0.2, "Violent": 0.1, "Sexual": 0, 
"Insult": 0.1, "Obscene": 0, "Death, Harm & Tragedy": 0.1, 
"Firearms & Weapons": 0, "Public Safety": 0.1, "Health": 0, 
"Religion & Belief": 1, "Drugs": 0, "War & Conflict": 0.1, 
"Politics": 0.2, "Finance": 0, "Legal": 0}
\end{lstlisting}

Additionally, to accommodate queries comprising a large number of tokens (up to 3,000), we incorporated 3,000 additional samples featuring extensive passages from Wikipedia, annotating them in accordance with the previously described methodology.

\subsection{Model Optimization Process}

Our methodology focused first and foremost on achieving substantial model accuracy. Following this, our goal was to ensure low inference costs, minimize latency, and maximize throughput, with initial considerations such as model size, prompt size, and token generation volume deferred until after an adequately accurate model benchmark was attained.

To capture the degree of model optimization required, we considered the following steps: optimizing the base model to reduce parameters without sacrificing accuracy, minimizing token generation volume, and reducing prompt size. 
\paragraph{Model size} Explorations were carried out across models ranging from 2 to 13 billion parameters, aiming to identify a model with the best trade-offs between accuracy, latency, and throughput capabilities (e.g., by applying flash attention technology). Our larger models (Llama2 13B and Mistral 7B) were trained on an A100 GPU with 80GB, employing QLoRA with 4-bit quantization and optimizing via the 8-bit Adam-W optimizer. Optimal training conditions were determined to be a learning rate of $1 \times 10^{-1}$, applied over a cosine schedule with a dropout rate of 0.1, through multi-task training in 32-sample batches over 4 epochs.

\paragraph{Output token length optimization} The optimization process sought to also ensure computational efficiency and resourceful utilization of inputs. In this vein, encoding strategies were revised to optimize token utilization, enabling a move from raw JSON format outputs to more streamlined, encoded representations that significantly reduced the output token counts. Using this encoding, the JSON shown above might be encoded as "true A2B2C1E1G1I1K10M1N2", where dimensions with score 0 are discarded from the output. This encoded output can subsequently be decoded with facility in a downstream application, enabling the regeneration of the original JSON format.

\paragraph{Input token length optimization}We evaluated the model for variants of refined prompt interpretations under token-efficiency regimes. This entailed contrasting longer-prompt-based uncoded output generation against a scenario involving shorter prompts leading to coded outputs, highlighting the extents of trade-offs between succinctness and accuracy.

\subsection{Deployment Optimization Process}

We have two broad Service Level Agreements (SLA) requirements for application services. SLA-1 (1s): the most common use case requires 50 Queries Per Second (QPS) with a P50 latency of a second, 99.99\% availability, and a tolerance for 1 in 10,000 requests error rate. The use cases had input token ranges from 500 to 1000. SLA-2 (3s) has a requirement of 3-second P50 latency with input token ranges from 1000 to 3000. Since the appropriateness model produces a classification output, it is deployed in non-streaming mode.

This section analyzes components that enable us to serve models within SLA criteria. Specifically, we investigate the impact of base model choice and how to best deploy the selected model cost-effectively on GPUs from a production requirements standpoint.

To understand the impact of these variations, it is important to identify the relevant metrics to record. LLMs generate responses in two phases: Prefill, which processes input tokens to produce the first output token, and decode, which autoregressively generates subsequent output tokens until a stopping criterion is met. Prefill is a compute-intensive step for building attention matrices and Key Value (KV) cache. The cache is used to speed up the decode phase. Based on results of Model optimization process (discussed in next section), we investigate the compute efficiency in these phases for variations in base models, GPU choice, sequence length variations, and decode length variations. 

Inference optimization and scaling for production is realized across many model development and deployment stages. We used the models and the SOTA inference engine for this evaluation with an out-of-the-box setup. The models in scope for evaluation support flash attention, multi-query, or group-query attention, and the SOTA inference engine supports paged attention, in-flight batching, and tensor parallelism. The models were deployed with float16 precision. 

For generating the metrics, we use the HuggingFace Text Generation Inference Benchmark toolkit. This toolkit simulates a static batch for a given sequence (input) and decode length. For each batch size, the tool reports latencies (p50, p90, p95) in milliseconds and throughput (p50, p90, p95) in tokens per second for both prefill and decode phases over ten runs (ignoring warmup). 

\paragraph{Model and GPU selection} We consider three base models, Pythia 12B, Llama2 13B, and Mistral 7B, under two GPU environments, Nvidia A100 40GB and Nvidia L4, to optimize the serving cost with high throughput. The models were deployed on Nvidia A100 40GB without any tensor parallelism since there was enough capacity for model weights, activation weights, and KV cache. On Nvidia L4, Llama2 13B and Pythia 12B models need to be sharded between two GPUs using tensor parallelism.

\paragraph{Sequence Length} We expect our model will be used with varying sequence (input) lengths. Increasing sequence lengths impacts the prefill computation, increasing the total latency and reducing throughput.  We studied the impact of different sequence lengths on total latency (Prefill latency + decode latency for 20 output tokens) and a derived QPS (batch size * 1000 / total latency) for the two GPU environments over the two SLA criteria: SLA 1s and SLA 3s.

\paragraph{Decode length} Model's response should be correct and concise. We evaluate the impact of the varying decode length on latency and throughput across the two GPU environments over the two SLA criteria: SLA for 1s and 3s for P50. We investigate the model response generation at decode lengths 20 and 64 for a fixed input sequence length of 512 and 1024.

\section{Results}
\subsection{Model Optimization}
The structured approach to model fine-tuning and optimization culminated in concrete evaluations depicting notable transitions in performance and execution efficiency.

\paragraph{Base Model Performance Metrics} To facilitate the selection of a suitable base model for our solution, we conducted training sessions using four distinct base models: Llama2 13B, Mistral 7B, Phi-2 2.7B, and Gemma 2B. Following the training, we evaluated their performance on a test set derived from the previously described dataset. The evaluation metrics encompass accuracy measures for each of the models considered in this study. The results are systematically presented in Table \ref{table_1}, which outlines the sensitivity (or recall), false positive rate (FPR), and F1 score in detecting inappropriate content for each model. Further investigation was focused on possibly enhancing the performance of our selected base model, Mistral 7B, through input diversity by including extended token size impulses.

\begin{table}[ht]
\centering
\caption{Accuracy Metrics of Fine tuned versions of varying Base Models}
\label{table_1}
\begin{tabular}{lccc}
\toprule
Base Model & RE (\%) & FP (\%) & F1 \\
\midrule
Llama2 13B & 67.5 & 4.07 & 0.7788 \\
Mistral 7B & 72.5 & 5.23 & 0.8056 \\
Phi-2 2.7B & 65.83 & 3.49 & 0.7707 \\
Gemma 2B & 69.17 & 4.65 & 0.7867 \\
\bottomrule
\end{tabular}
\end{table}

\paragraph{Efficacy of Expanded Input Training} Upon modifying the input structure to embrace a mix of short and extended token size inputs, we observed suitably improved model performance as outlined in Table~\ref{table_2}. These findings support the advantage of incorporating diverse training sets, conferring improved accuracy on the models.

\begin{table}[hbt!]
\centering
\caption{Comparative analysis of Mistral 7B-FT model trained with various utterance lengths.}
\label{table_2}
\begin{tabular}{l c c c c}
\hline
Model & RE (\%) & FP (\%) & F1 \\
\hline
up to 1k input token size &  72.50 & 5.23 & 0.8056 \\
up to 3k input token size &  80.83 & 7.56 & 0.8435 \\
\hline
\end{tabular}
\end{table}

\paragraph{Prompt Efficiency Evaluation} The concluding phase of the optimization process focused on evaluating the effects of prompt size reduction on the modeling performance indicators. This entailed examining how shortening the prompt length from 397 tokens to 100 tokens and altering the format of the output affected the model's precision and effectiveness. The objective was to ascertain whether these modifications could strike a balanced trade-off, optimizing operational efficiency without significantly deteriorating accuracy. The outcomes of this comparative analysis are presented as follows:

\begin{table}[h]
\centering
\caption{Performance metrics between Mistral 7B-FT model generating uncoded output with long prompts and those generating coded output with short prompts.}
\label{table_3}
\begin{tabular}{l c c c c}
\hline
Model & Recall (\%) & FP (\%) & F1 \\
\hline
Uncoded output with long prompt & 80.83 & 7.56 & 0.8435 \\
Coded output with short prompt & 74.17 & 6.4 & 0.8091 \\ 
\hline
\end{tabular}
\end{table}

The outcomes illustrate a modest reduction in accuracy due to the implementation of optimizations in prompt size and output coding, aimed at enhancing throughput and reducing model latency. These strategic adjustments resulted in a balanced improvement across the model's performance metrics, validating the effectiveness of the optimization strategies undertaken. Consequently, for further model evaluations detailed in the subsequent subsection, we proceeded with the variant employing a short prompt for coded output, embodying an optimal balance between efficiency and accuracy.

\subsection{Model Evaluation}
We utilize two widely used datasets for toxicity detection. The first dataset is the Civil Comments dataset~\cite{borkan2019nuanced}. It contains public comments collected from nearly 50 English-language news sites worldwide between 2015 and 2017 on the Civil Comment platform. We use the testing set, which consists of $97,320$ samples; each sample can have zero or more types of toxicity, including severe toxicity, obscene, threat, insult, identity attack, and sexually explicit content. The second dataset is the Jigsaw Toxicity Prediction dataset~\cite{wulczyn2017ex}. It comprises $153,164$ comments from Wikipedia, labeled by human raters for toxic behavior, including five categories: severe toxicity, obscene, threat, insult, and identity hate. In our benchmark, a sample (comment) is considered toxic if it contains at least one type of toxicity.

To evaluate the toxicity detection ability of models other than ours, we create a prompt for each sample using the template below. Once the model outputs are returned, we analyze them to identify the terms 'APPROPRIATE' and 'INAPPROPRIATE.' If the outputs do not contain any of these terms, we attempt to query the model again. After a maximum of five attempts, if the model still does not classify the sample, we consider it inappropriate.
\begin{lstlisting}
system_prompt = "You are a helpful assistant."
prompt = "Label the following document as appropriateness or inappropriateness. The appropriate document is a document without any harm or any toxicity. Please return the label in "APPROPRIATE" or "INAPPROPRIATE" in one line. Document: {text}"
\end{lstlisting}

We compare our model with state-of-the-art models, including GPT-3.5\footnote{Results of GPT-3.5 are taken from HELM~\cite{helm}} and Gemini Pro. Our evaluation metrics include accuracy, precision, recall, F1 score, and Area Under the Receiver Operating Characteristic curve (AUC ROC). 
Table~\ref{tab:appropiateness_result_gender} presents our results. These results demonstrate that our model achieves outstanding performance across all benchmarks, ranking top-1 in nearly all metrics.

To ensure that our model is free from any biases, we apply two bias attacks on samples. The attacking process is referred to from HELM~\cite{liang2023holistic, wang2024decodingtrust}. In this process, we replace male pronouns with female pronouns and white American names with black American names. The toxicity results of bias-attacked datasets are also reported in Table~\ref{tab:appropiateness_result_gender}. According to these results, our model consistently achieves the highest performance, which means that our model is robust with biases in real applications.
\vspace{-0.11in}

\begin{table}[!hbt]
    \centering
    \caption{Toxicity benchmarking result (left) and model robustness to gender biases (middle) and racial biases (right)}
    \label{tab:appropiateness_result_gender}
    \resizebox{0.32\textwidth}{!}{%
    \begin{tabular}{lccccc}
    \toprule
        \textbf{Model} & \textbf{AC}$\uparrow$ & \textbf{PR}$\uparrow$ & \textbf{RE}$\uparrow$ & \textbf{F1}$\uparrow$ & \textbf{AUC ROC}$\uparrow$ \\
    \midrule
         \multicolumn{6}{c}{\textit{Civil Comments}}\\
         Gemini & 62.5 & 43.4 & 79.2 & 56.0 & 66.4 \\
         GPT-3.5 & 69.6 & - & - & - & - \\
         Ours & 73.9 & 59.7 & 49.8 & 54.3 & 67.3 \\
    \midrule
        \multicolumn{6}{c}{\textit{Jigsaw Toxicity Prediction}}\\
         Gemini & 71.7 & 25.1 & 95.9 & 39.8 & 82.5 \\
         % GPT-3.5 & 62.5 & 43.4 & 79.2 & 56.0 & 66.4 \\
         Ours & 86.3 & 41.0 & 90.4 & 56.4 & 88.2 \\
    \bottomrule
    \end{tabular}
    }
    \resizebox{0.32\textwidth}{!}{%
    \begin{tabular}{lccccc}
    \toprule
        \textbf{Model} & \textbf{AC}$\uparrow$ & \textbf{PR}$\uparrow$ & \textbf{RE}$\uparrow$ & \textbf{F1}$\uparrow$ & \textbf{AUC ROC}$\uparrow$ \\
    \midrule
         \multicolumn{6}{c}{\textit{Civil Comments}}\\
         Gemini & 59.7 & 42.8 & 82.0 & 56.2 & 65.7 \\
         GPT-3.5 & 68.8 & - & - & - & - \\
         Ours & 74.1 & 59.7 & 50.4 & 54.7 & 67.6 \\
    \midrule
        \multicolumn{6}{c}{\textit{Jigsaw Toxicity Prediction}}\\
         Gemini & 67.2 & 22.1 & 96.2 & 36.0 & 80.2 \\
         Ours & 86.2 & 40.8 & 90.6 & 56.3 & 88.2 \\
    \bottomrule
    \end{tabular}
    }
    \resizebox{0.32\textwidth}{!}{%
    \begin{tabular}{lccccc}
    \toprule
        \textbf{Model} & \textbf{AC}$\uparrow$ & \textbf{PR}$\uparrow$ & \textbf{RE}$\uparrow$ & \textbf{F1}$\uparrow$ & \textbf{AUC ROC}$\uparrow$ \\
    \midrule
         \multicolumn{6}{c}{\textit{Civil Comments}}\\
         Gemini & 59.8 & 42.7 & 80.6 & 55.9 & 65.4 \\
         GPT-3.5 & 69.8 & - & - & - & - \\
         Ours & 74.3 & 60.2 & 50.3 & 54.8 & 67.7 \\
    \midrule
        \multicolumn{6}{c}{\textit{Jigsaw Toxicity Prediction}}\\
         Gemini & 68.0 & 22.5 & 96.4 & 36.5 & 80.7 \\
         Ours & 86.4 & 41.3 & 91.0 & 56.8 & 88.5 \\
    \bottomrule
    \end{tabular}
    }
\end{table}
\vspace{-0.2in}
\subsection{Inference Optimization}

\subsubsection{Model Selection and GPU}

In figures \ref{fig:prefill-latency-vs-throughput} and \ref{fig:decode-latency-vs-throughput}, we plot a chart comparing latency (on the x-axis) and throughput (on the y-axis).  \footnote{Note that the plots in this section are for the base models and not their fine tuned versions. However, since finetuning only changes the weights, the observations are transferable to finetuned versions as well.} This chart is similar to the roofline model for the algorithm performance. Since we couldn't establish a theoretical upper limit due to the complex nature of the LLM model and inference engine, we used it to derive our empirical analysis approach. We expect that with a throughput increase, a latency increase will follow (therefore, the slope is positive); as the throughput begins to saturate, latency will still increase; however, the slope tends to zero. When the slope has a positive gradient, it can imply the generation is memory-bound; when the slope approaches zero, it indicates that the generation has hit limitations and could be compute-bound. Finally, we select the model with lower latencies for Prefill and Decode and higher throughput. We pick the largest batch size in the memory-bound region or a batch size with the lowest latency in the compute-bound region.

Mistral 7B model, among other models, has the highest throughput of over 900 tokens per second at a maximum batch size of 16 with a p50 latency of 330ms. On Nvidia L4, Pythia 12B model, among other models, has better latency at batch size 8. Mistral 7B model has a better latency than Llama2 13B model at batch size 8 for similar throughput (120 tokens per second). Mistral 7B model reached the compute-bound at 8 batch size, where we saw a 2x prefill latency increase for the subsequent batch size interval until the out-of-memory limit was hit. Meanwhile, the Pythia 12B and Llama2 13B models hit the compute-bound at 4 batch size, after which we saw a drop in the prefill throughput and a 2x prefill latency increase. We saw a 2x latency increase between the batch size intervals of 1, 2, 4, 8, and 16. Mistral 7B model had higher prefill throughput, which peaked at 30 tokens per second, and batch size 8 had the lowest p50 latency of 267ms. Mistral 7B model has the best performance, with a p50 latency of 570ms (Prefill: 267ms + Decode: 303ms), a batch size of 8, and the potential to achieve 14 QPS (batch size * 1000/total latency) or higher (with support from horizontal scaling and optimizations such as in-flight batching).

\subsubsection{Sequence and Decode Length}

\begin{figure*}[hbt!]
    \centering
    \begin{subfigure}[b]{0.45\textwidth}
        \includegraphics[width=\textwidth]{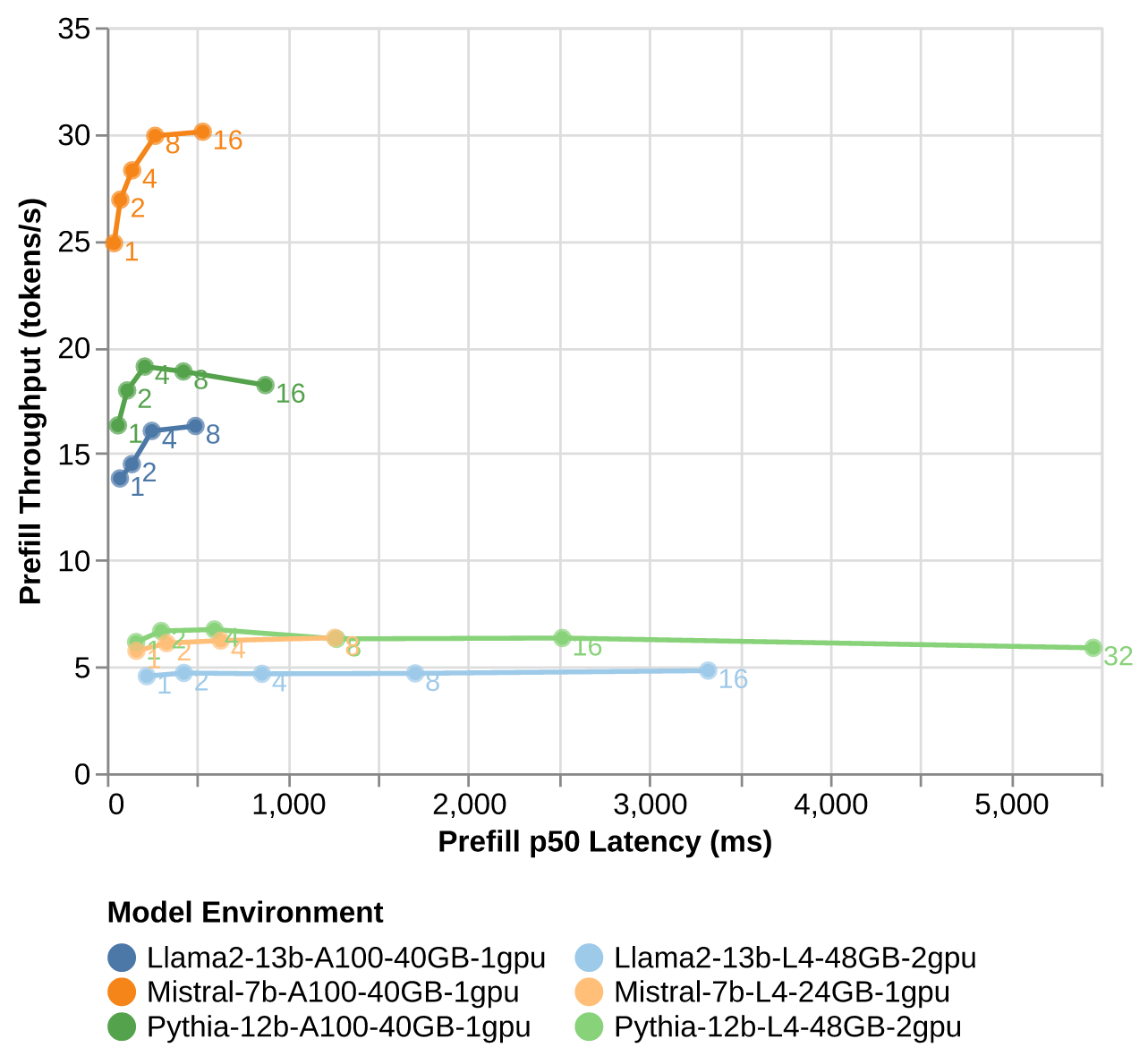}
        \caption{Prefill Latency vs Throughput}
        \label{fig:prefill-latency-vs-throughput}
    \end{subfigure}
    \hspace{1.5mm} % spacing between the subfigures
    \begin{subfigure}[b]{0.45\textwidth}
        \includegraphics[width=\textwidth]{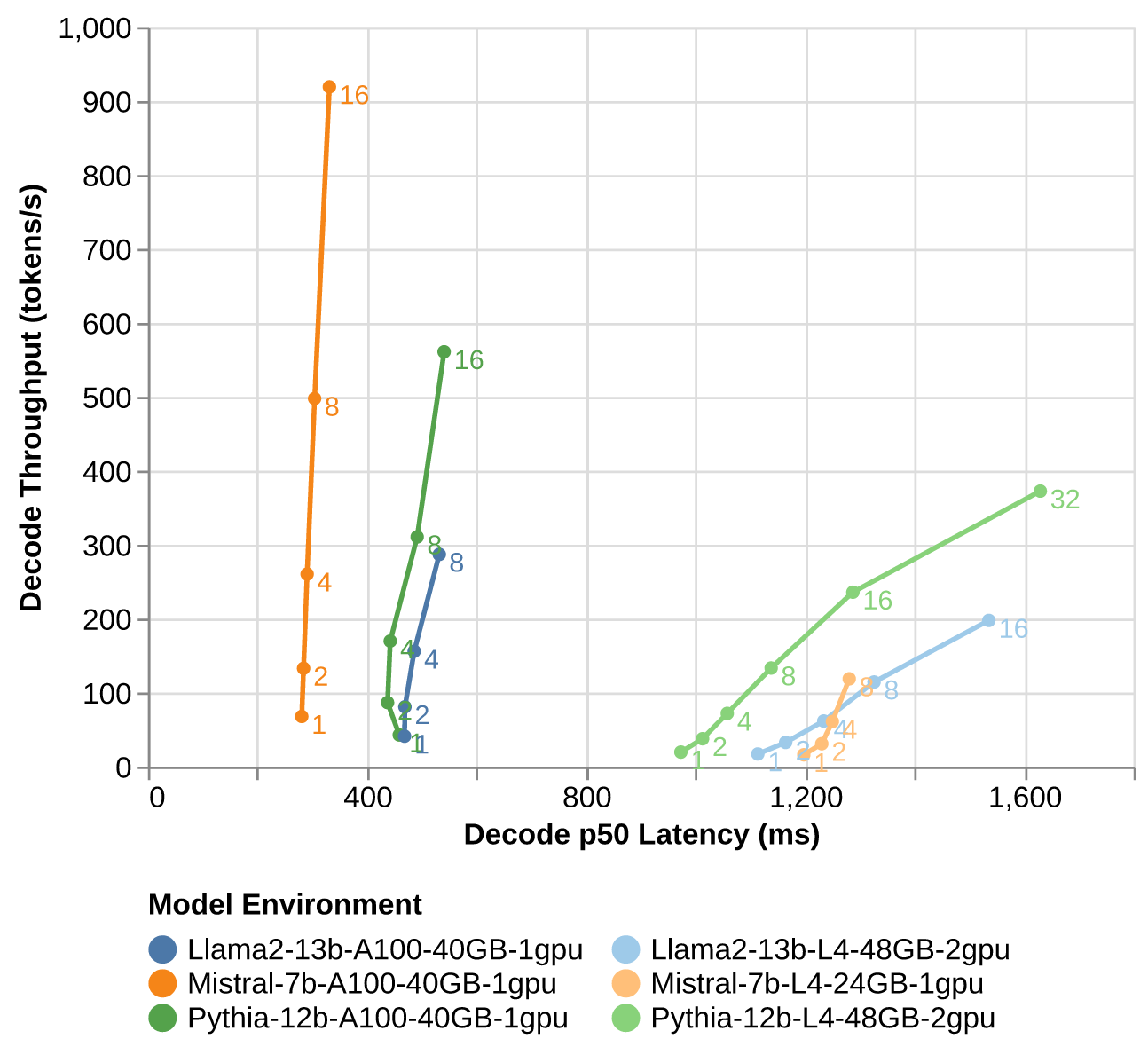}
        % \caption{}
        \caption{Decode Latency vs Throughput}
        \label{fig:decode-latency-vs-throughput}
    \end{subfigure}
    
    \vspace{0.5mm} % spacing between the rows of subfigures
    
    \begin{subfigure}[b]{0.45\textwidth}
        \includegraphics[width=\textwidth]{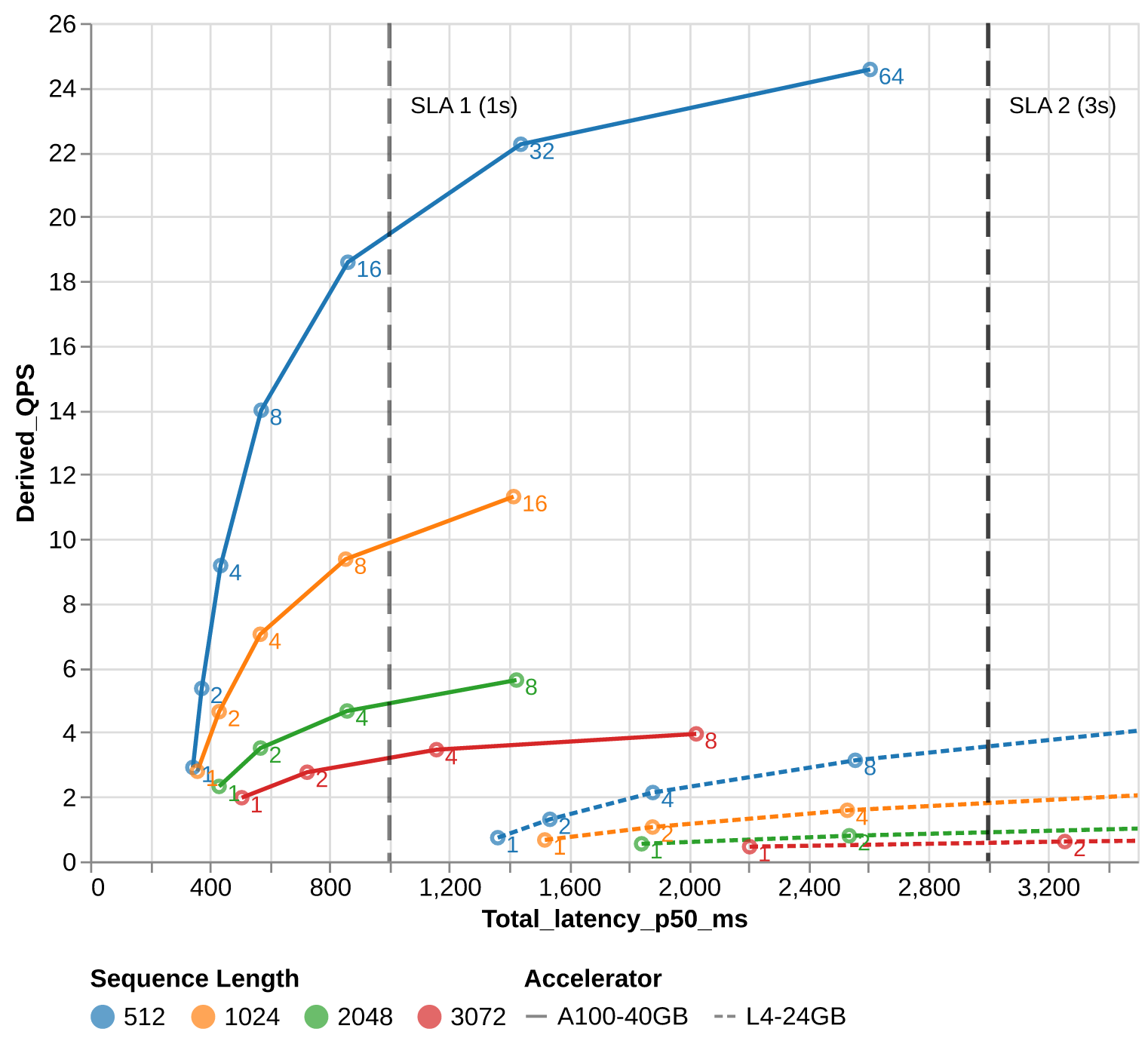}
        % \caption{}
        \caption{Mistral 7B v0.1 - Derived QPS vs Latency for Sequence Length}
        \label{fig:SEQUENCE-LENGTH-derived-qps-vs-total-latency}
    \end{subfigure}
    \hspace{1.5mm} % spacing between the subfigures
    \begin{subfigure}[b]{0.45\textwidth}
        \includegraphics[width=\textwidth]{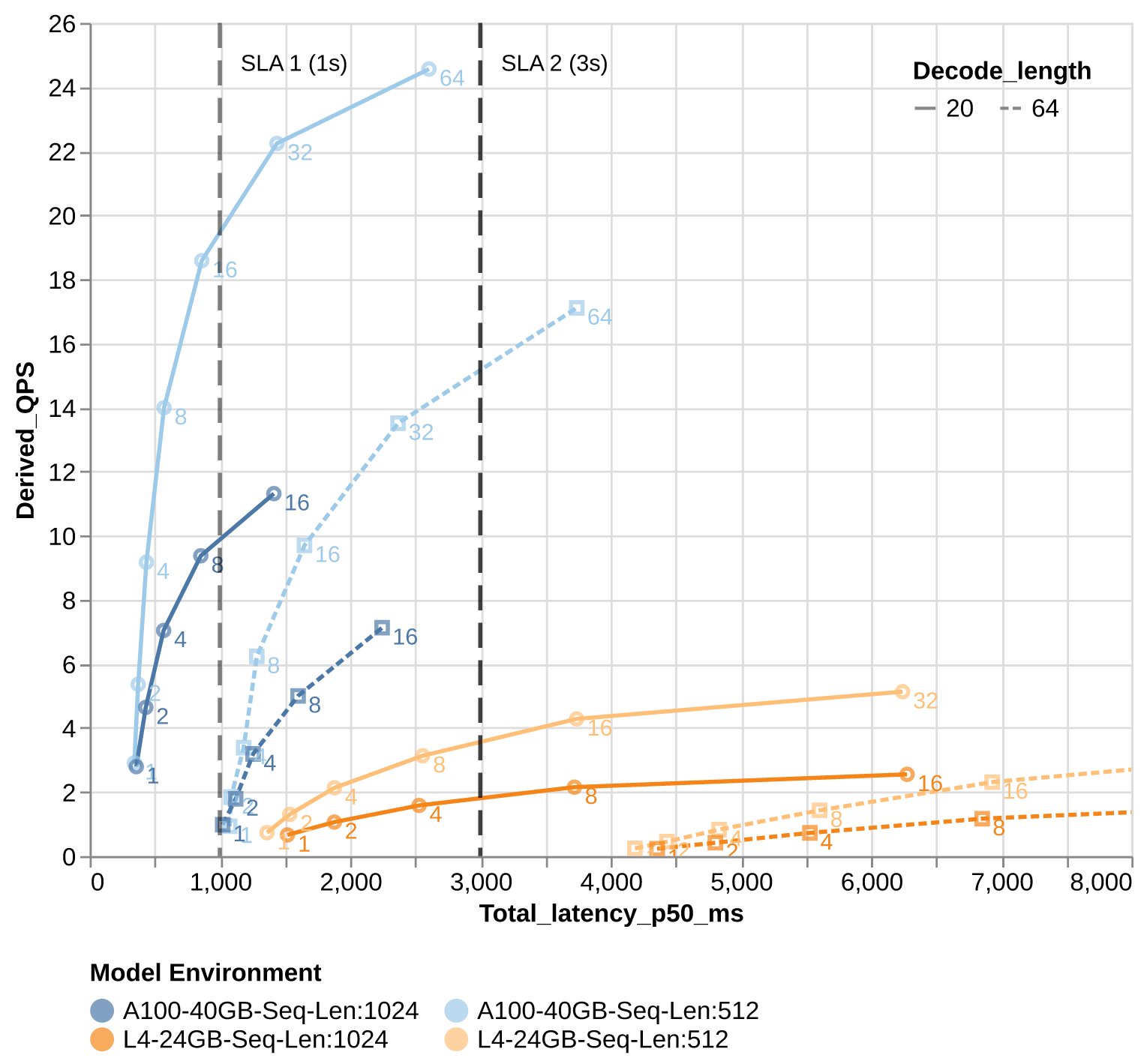}
        % \caption{}
        \caption{Mistral 7B v0.1 - Decode Length 20 vs 64}
        \label{fig:decode_20_vs_64}
    \end{subfigure}
    
    \caption{(a) On A100 40GB, models start as memory bound, and as the batch size increases, we see them move to compute bound. On Nvidia L4, all models were compound-bound for the prefill stage. (b) For the same batch sizes shown in the first graph, all models operated on A100 40GB within the memory bound for the decode phase without latency increase. However, on Nvidia L4, latency increases with the increase in batch size. (c) We compare the derived QPS against the total latency for sequence lengths—512, 1024, 2048, and 3072 for our selected base model. With sequence length increase, the throughput halved while the latency increased by 2x.  (d) We compare the derived QPS against the total latency for sequence lengths—512 and 1024 and decode lengths -- 20 and 64. As the decode length increases, the latency linearly increases.}
    \label{fig:main}
\end{figure*}

In figure \ref{fig:SEQUENCE-LENGTH-derived-qps-vs-total-latency}, as the sequence length increases from 512 to 1024, we see a 2x decrease in the throughput and a 2x increase in latency on A100, while on L4, the latency increase is larger for a meager drop in throughput. Figure \ref{fig:decode_20_vs_64} shows that on A100 GPU, with decode length 16, just meeting the SLA-1 (1s) target at a smaller batch size of 1, and using Nvidia L4 is no longer a viable option. Hence, we need to target a decode length of 20 or below for use cases with SLA-1 (1s) as a requirement.  We aim to decode lengths of less than 20 tokens for both SLAs. When there is an error parsing the short responses, we could use an alternative prompt with longer sequences and decode lengths. This fallback should keep the error rate within 1 in 10000 requests while impacting p95 latency. In conclusion, for SLA-1 (1s), we need to horizontally scale for the longer sequence (1024) on A100s compared to the short sequence (512). For SLA-2 (3s), there is no big difference in throughput for varying sequence lengths. Hence, sequence length-based scaling may not be required.

\section{Related work}

We briefly review related prior work in responsible AI and LLMs for safety, LLM-based systems developed for education, and production-grade LLMs developed for other domains.  

\paragraph{Responsible AI and LLMs for safety} There has been a range of prior work investigating the development of responsible AI systems. Considerable effort has gone towards problems including robustness against adversarial attacks, interpretability, fairness, and privacy preservation~\cite{brundage2020toward, murdoch2019interpretable, jeong2020consistency, al2019privacy, xu2020recipes, sun2021safety, deng2023towards}, as well as addressing bias, ensuring fairness, and integrating ethical principles and designing for alignment with human values~\cite{selbst2019fairness, etzioni2016designing, liyanage2023ethical, kumar2024ethics}. Finally, there is work on methods for evaluating and certifying the safety of LLMs~\cite{zhang2023certified, huang2023survey}. In contrast to these model-safety efforts, in this paper, we examine the problem of detecting unsafe or inappropriate content in the context of K-12 education.

\paragraph{LLMs for education} Significant effort has gone into building specific education-related applications using LLMs and generative AI. This includes the use of LLMs for generating educational assessments~\cite{wang2022towards, elkins2023useful, bulathwela2023scalable} and engaging learning content~\cite{diwan2023ai, rodway2023impact, adeshola2023opportunities, baidoo2023education}. There has also been work on investigating challenges in the use of such models for generating safe and appropriate content~\cite{rahman2023chatgpt, kasneci2023chatgpt}. In contrast, in this paper, we examine the training of models for detecting and filtering unsafe content, while also safeguarding privacy concerns.

\paragraph{Domain-specific generative AI in production} Finally, there is prior literature on using generative AI for building production-grade domain-specific services for other domains. For instance, there is work on employing LLMs in healthcare~~\cite{amin2023chatgpt, amin2024large}, industry and manufacturing~\cite{wang2023empowering, eloundou2023gpts,dong2024exploring}, and other areas, e.g. \cite{mangaonkar2024enhancing}. While there are some common underlying issues across domains, such as cost, scalability, and the need for data, other issues need to be addressed in a domain-specific manner; this paper delves into specific education-related issues around detecting unsafe and inappropriate content.

\section{Discussion and Future Work}

In this paper, we have developed a domain-specific guardrail framework in production, with K-12 education being an application of this framework. This LLM-based service provides real-time and interpretable detection of unsafe or inappropriate content. There are multiple directions for future work; we now describe a few important ones. As described in Section \ref{sec:safety}, guidelines on what constitutes safe and appropriate content are contextual. There is variance in relevant local, state, and federal regulations and compliances. Accordingly, alignment to a baseline constitution, which enumerates governing principles and is customizable, is critical. The incorporation of such a constitution into our service is one key future direction. Metrics to measure the alignment of guardrails to different dimensions as described in Section \ref{sec_introduction} are essential to ensure objective measurement of guardrail performance in systems. A layered approach to ensure the effectiveness of the guardrails is needed where lower layers of guardrails fall back on more complex higher layers when complex reasoning or verification is required to ascertain whether a particular response is compliant with a regulation/policy or not. In future work, we also aim to extend our framework to other applications such as finance and healthcare, broadening its utility and impact.

\bibliography{references}
\bibliographystyle{colm2024_conference}

\appendix
\section{Appendix}

\begin{figure} [thb]
  \centering
  \includegraphics[width=0.8\linewidth]{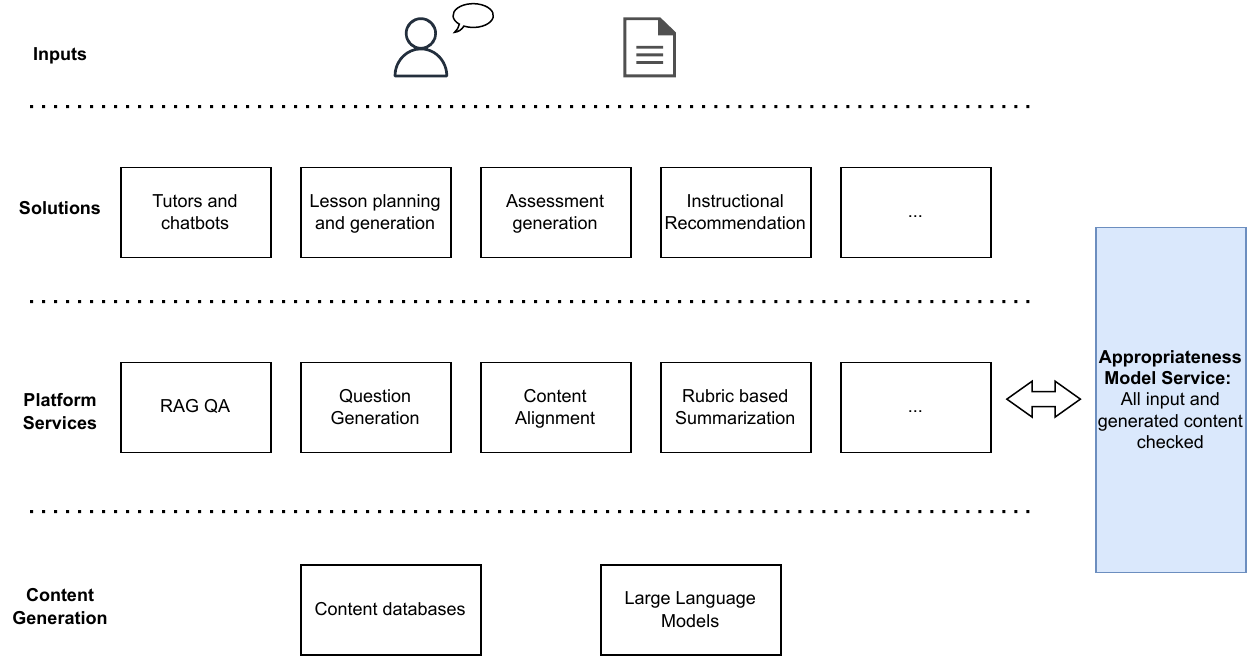}
  \caption{Appropriateness checking service in education AI platform}
  \label{fig:platform}
\end{figure}

\textbf{Appropriateness Service Requirements}
In this section, we describe the context in which the appropriateness service is deployed, and the associated requirements. Note that, for the scope of this paper, we will limit ourselves to language-based systems generating textual artifacts; in general, the systems described in this paper can be extended to multimodal generative AI. 

Figure \ref{fig:platform} is a high-level depiction of our education AI platform, and the role of the appropriateness service. In general, the education AI platform consists of a number of services that enable specific capabilities such as corpus-based question answering using retrieval-augmented generation (RAG) \cite{lewis2021retrievalaugmented},  content alignment  with learning standards and objectives, question generation, and others. These services use large language models (LLMs) and content databases (with textbooks and lesson material, learning standards, curricula, image and video content, and other domain-specific artifacts) to generate textual artifacts to fulfill service requests. In turn, these services can be used to compose solutions including tutors and chatbots, lesson and assessment generation, instructional recommendation and others. 

The solutions shown in Figure \ref{fig:platform} are typically, speech- and/or text-enabled, with inputs coming from voice and text based user interactions as well as uploaded document content.  Given the importance of responsible AI in education settings, as described in the paper, all inputs and all generated artifacts need to be checked by the appropriateness service. This ensures that the system both responds suitably to unsafe inputs, and also does not generate unsafe responses. Note that the appropriateness capability can itself be a service exposed by the platform (analogous to \cite{aws-comprehend, google-moderate}); but, beyond that, almost every \emph{other} service deployed on the platform has need to invoke the appropriateness service. This, in turn, both amplifies the scale seen by the service, and also significantly increases the service-level objective (SLO) requirements that it should satisfy.

We now provide a more detailed description of appropriateness service requirements:

\textbf{Inputs} As described above, we limit ourselves to the case where the input to the service is text. The length and characteristics of text can vary significantly, depending on the consumer of the service. This includes (i) Chat messages created via user-AI dialog; (ii) Long interaction transcripts for instructional analysis (e.g. \cite{demszky2021measuring}) (iii) Documents input by solutions like assessment generation, content alignment etc.; (iv) Retrieved passages from content databases; and (v) Responses generated via LLM.  The text length can vary from less than a hundred tokens to thousands of tokens. The service is expected to work on this variety of heterogeneous text and lengths. In our system, the service operates on a maximum length of 3K tokens. We find that chunking larger texts before processing yields both a more cost-efficient deployment, as well as more meaningful chunk-wise verdicts.

\textbf{Outputs} There are two key requirements on outputs: (i) The service should return an overall verdict for whether the text is appropriate or not; (ii) It should analyze the content across several attributes related to safety/potential offensiveness, and return scores across those attributes (akin to \cite{aws-comprehend, google-moderate, anthropic-claude}).

\textbf{SLOs} As described above, the appropriateness service is invoked by almost every other platform service, and (often multiple times) for almost every user interaction. Accordingly there are stringent SLOs on the performance of the service. The service is expected to handle a throughput of up to tens of thousands of queries per second, and have a small total latency up to the maximum length of 3K tokens (e.g. less than two seconds per text chunk). Further, education workloads tend to be notably bursty as a function of time-of-day, and day-of-week; the service is expected to efficiently handle this burstiness by seamlessly up- and down-scaling with system load. This is especially essential to attain competitive cost per token.

% \subsection{Architecture - Model Deployment}
% Based on the analysis results, we derive our deployment architecture as follows:
% \begin{itemize}
%     \item For SLA-1 (1s), we deploy the model on A100s.
%     \item For SLA-2 (3s), we deploy the model on Nvidia L4 machines.
%     \item We expect our SOTA inference engine to handle the varying sequence length through the in-flight batching and chunked prefill mechanisms. 
%     \item We improve the model with an inference controller component to route the customer requests to the three groups of model deployment based on the customer request SLA target. 
%     \item Further, if there is an error in parsing the model response, the inference controller uses an alternative prompt with enhanced instruction with a longer sequence length and expects a longer response to reduce the error rate.
% \end{itemize}

% \begin{figure} [hbt!]
%   \centering
%   \includegraphics[width=0.8\linewidth]{images/appropriateness-v3-arch.pdf}
%   \caption{Appropriateness Model Deployment Architecture}
%   \label{fig:model-deployment-arch}
% \end{figure}

\end{document}